\documentstyle[12pt]{article}

\catcode`\@=11
\long\def\@makefntext#1{
\protect\noindent \hbox to 3.2pt {\hskip-.9pt
$^{{\ninerm\@thefnmark}}$\hfil}#1\hfill}                

\def\@makefnmark{\hbox to 0pt{$^{\@thefnmark}$\hss}}  

\def\ps@myheadings{\let\@mkboth\@gobbletwo
\def\@oddhead{\hbox{}
\rightmark\hfil\ninerm\thepage}
\def\@oddfoot{}\def\@evenhead{\ninerm\thepage\hfil
\leftmark\hbox{}}\def\@evenfoot{}
\def\sectionmark##1{}\def\subsectionmark##1{}}

\renewcommand{\thefootnote}{\fnsymbol{footnote}}

\newcounter{sectionc}\newcounter{subsectionc}\newcounter{subsubsectionc}
\renewcommand{\section}[1] {\vspace*{0.6cm}\addtocounter{sectionc}{1}
\setcounter{subsectionc}{0}\setcounter{subsubsectionc}{0}\noindent
	{\normalsize\bf\thesectionc. #1}\par\vspace*{0.4cm}}
\renewcommand{\subsection}[1] {\vspace*{0.6cm}\addtocounter{subsectionc}{1}
	\setcounter{subsubsectionc}{0}\noindent
	{\normalsize\it\thesectionc.\thesubsectionc. #1}\par\vspace*{0.4cm}}
\renewcommand{\subsubsection}[1]
{\vspace*{0.6cm}\addtocounter{subsubsectionc}{1}
	\noindent {\normalsize\rm\thesectionc.\thesubsectionc.\thesubsubsectionc.
	#1}\par\vspace*{0.4cm}}

\newcounter{appendixc}
\newcounter{subappendixc}[appendixc]
\newcounter{subsubappendixc}[subappendixc]

\renewcommand{\appendix}[1] {\vspace*{0.6cm}
	\refstepcounter{appendixc}
	\setcounter{figure}{0}
	\setcounter{table}{0}
	\setcounter{equation}{0}
	\renewcommand{\thefigure}{\Alph{appendixc}.\arabic{figure}}
	\renewcommand{\thetable}{\Alph{appendixc}.\arabic{table}}
	\renewcommand{\theappendixc}{\Alph{appendixc}}
	\renewcommand{\theequation}{\Alph{appendixc}.\arabic{equation}}
	\noindent{\bf Appendix \theappendixc #1}\par\vspace*{0.4cm}}

\def\abstracts#1{{

\centering{\begin{minipage}{12.2truecm}\footnotesize\baselineskip=12pt\noindent
	\centerline{\footnotesize ABSTRACT}\vspace*{0.3cm}
	\parindent=0pt #1
	\end{minipage}}\par}}

\renewenvironment{thebibliography}[1]
	{\begin{list}{\arabic{enumi}.}
	{\usecounter{enumi}\setlength{\parsep}{0pt}
\setlength{\leftmargin 1.25cm}{\rightmargin 0pt}
	 \setlength{\itemsep}{0pt} \settowidth
	{\labelwidth}{#1.}\sloppy}}{\end{list}}

\newcounter{itemlistc}
\newcounter{romanlistc}
\newcounter{alphlistc}
\newcounter{arabiclistc}

\newcommand{\fcaption}[1]{
	\refstepcounter{figure}
	\setbox\@tempboxa = \hbox{\footnotesize Fig.~\thefigure. #1}
	\ifdim \wd\@tempboxa > 6in
	   {\begin{center}
	\parbox{6in}{\footnotesize\baselineskip=12pt Fig.~\thefigure. #1}
	    \end{center}}
	\else
	     {\begin{center}
	     {\footnotesize Fig.~\thefigure. #1}
	      \end{center}}
	\fi}

\newcommand{\tcaption}[1]{
	\refstepcounter{table}
	\setbox\@tempboxa = \hbox{\footnotesize Table~\thetable. #1}
	\ifdim \wd\@tempboxa > 6in
	   {\begin{center}
	\parbox{6in}{\footnotesize\baselineskip=12pt Table~\thetable. #1}
	    \end{center}}
	\else
	     {\begin{center}
	     {\footnotesize Table~\thetable. #1}
	      \end{center}}
	\fi}

\def\@citex[#1]#2{\if@filesw\immediate\write\@auxout
	{\string\citation{#2}}\fi
\def\@citea{}\@cite{\@for\@citeb:=#2\do
	{\@citea\def\@citea{,}\@ifundefined
	{b@\@citeb}{{\bf ?}\@warning
	{Citation `\@citeb' on page \thepage \space undefined}}
	{\csname b@\@citeb\endcsname}}}{#1}}

\newif\if@cghi
\def\cite{\@cghitrue\@ifnextchar [{\@tempswatrue
	\@citex}{\@tempswafalse\@citex[]}}
\def\citelow{\@cghifalse\@ifnextchar [{\@tempswatrue
	\@citex}{\@tempswafalse\@citex[]}}
\def\@cite#1#2{{$\null^{#1}$\if@tempswa\typeout
	{IJCGA warning: optional citation argument
	ignored: `#2'} \fi}}

 1
 1
 1

\font\ninerm=cmr9

\begin{document}

\centerline{\normalsize\bf $b\to s \gamma$ DECAY
 IN THE TWO HIGGS DOUBLET MODEL
\footnote{This work is partly supported by China Postdoctoral Science
Foundation.}}

\baselineskip=16pt

\vspace*{0.6cm}
\centerline{\footnotesize Cai-Dian L\"u}
\baselineskip=13pt
\centerline{\footnotesize\it  CCAST (World Laboratory), P.O. Box 8730,
Beijing 100080, China}
\baselineskip=12pt
\centerline{\footnotesize\it Institute of Theoretical Physics,
P.O. Box 2735, Beijing 100080, China}
\centerline{\footnotesize E-mail: lucd@itp.ac.cn}
\baselineskip=13pt

\vspace*{0.9cm}
\abstracts{ QCD corrections to $b \to s \gamma$  decay in the two
Higgs doublet model are calculated from
the energy scale of top quark to that of bottom.
The constraints on the two Higgs doublet model
from the new experimental bounds of $b\to s\gamma$ by CLEO and the
latest top quark mass by CDF and D0 are reanalyzed.
It shows that the constraints become more
stringent than that of the earlier analysis,
i.e. a bigger region of the parameter space of the model is ruled out.}

\vspace*{0.6cm}
\normalsize\baselineskip=15pt
\setcounter{footnote}{0}
\renewcommand{\thefootnote}{\alph{footnote}}

It is known that the experimental bounds of $b\to
s\gamma$ set very strong constraints on the two Higgs doublet
model (2HDM), a minimal extension of the Standard Model (SM).
In additional to searching for the neutral Higgs of minimal SM,
phenomenologically to investigate
possible extensions of SM is also another hot topic in particle
physics, thus to apply the latest experimental results of the measurement
on $b\to s\gamma$~\cite{cleo2} and the newly discovery of top quark~\cite{t1}
to reexamine the constraints on the 2HDM so as to upgrade
the allowed values of the model parameters is no doubt always
to be interesting.

Reviewing the earlier analysis~\cite{s1}, one would find that
the QCD correction effects owing to the change of the energy scale
from top quark's down to that of $W$ boson were ignored.
Indeed this piece of QCD correction itself is not great, but we
treat them seriously~\cite{lcd1}, and finally find it being not negligible
since this correction affects the constraints on the two
Higgs doublet model sizable in the report.

There are two ways for 2HDM
to avoid tree-level flavor changing neutral currents (FCNCs).
The first (Model I) is to allow only one of the two Higgs doublets to
couple to both types, u-type and d-type, of quarks~\cite{Hab} but
the other doublet is totally forbidden by certain discrete symmetry.
The second (Model II) is to arrange as that
one Higgs doublet couples to u-type quarks while the other
couples to d-type quarks respectively due to a different
discrete symmetry~\cite{Gla}.
It is of interest to note that the Model II,
as a natural feature, occurs in such a theory as that with
supersymmetry or with a Peccei-Quinn type of symmetry.

The effective Hamiltonian after integrating out the heavy top
freedom is:
\begin{equation}
{\cal H}_{eff}=2 \sqrt{2} G_F V_{tb}V_{ts}^*\displaystyle \sum _i
C_i(\mu)O_i(\mu). \label{eff}
\end{equation}
The coefficients $C_i(m_t )$ of effective operators $O_i$
can be calculated from matching conditions at $\mu=m_t$~\cite{lcd1},
 and $C_i(\mu )$ can be obtained from
their renormalization group equation(RGE):
\begin{equation}
\mu \frac{d}{d\mu} C_i(\mu)=\displaystyle\sum_{j}(\gamma^{\tau})_{
ij}C_j(\mu),\label{ren}
\end{equation}
This is the very procedure which gives out the QCD corrections from
$m_t$ scale to $M_W$ scale. In this stage, we calculate the anomalous
dimensions $\gamma_{ij}$
of operators~\cite{lcd2}, it shows that there are errors
in the previous calculations~\cite{cho}. After corrections, we
got the coefficients of operators at $\mu=M_W$ by
renormalization  group running. The coefficients of operators at
this stage changed due to this corrections, especially
 for operator $O_8$~\cite{lcd1}.
When $\mu = M_W$, one encounters with the
W boson. If integrating out the $W$ boson freedom further, once more
six relevant four-quark operators will be added~\cite{lcd1}.
Thus one obtains the effective Hamiltonian just below $M_W$ scale.

The running of the coefficients of operators from $\mu=M_W$ to $\mu=m_b$
was well described in refs.~\cite{Grin}. With the
running due to QCD, the coefficient of the operator at $\mu=m_b$ scale is:
\begin{equation}
C_7^{eff}(m_b) = \eta^{16/23}C_7(M_W) +\frac{8}{3}
( \eta^{14/23}-\eta^{16/23} ) C_8(M_W)
+C_2(M_W) \displaystyle \sum _{i=1}^{8} h_i \eta^{a_i}.
\end{equation}
Here $\eta = \alpha_s(M_W) /\alpha_s (m_b)$,
$$ h_i =\left( \frac{626126}{272277}, -\frac{56281}{51730},
-\frac{3}{7}, -\frac{1}{14}, -0.6494, -0.0380, -0.0186, -0.0057 \right),$$
$$a_i = \left( \frac{14}{23}, \frac{16}{23}, \frac{6}{23}, -\frac{12}{23},
0.4086, -0.4230, -0.8994, 0.1456 \right).$$
The explicit expressions of
the coefficient of operators at $\mu=M_{W}$ are given at previous
paper~\cite{lcd1},

Using the quite well established
semileptonic decay rate $Br(B \to X_c e\overline{\nu} )$,
one obtains~\cite{Grin},
\begin{eqnarray}
\frac{BR(B \rightarrow X_s \gamma)}{BR(B
\rightarrow X_c e \overline{\nu})} \simeq \frac{|V_{ts}^*V_{tb}|^2}{
|V_{cb}|^2} \frac{6 \alpha_{QED}}{\pi g (m_c/m_b)}
|C_7^{eff}(m_b)|^2,
\end{eqnarray}
where the phase space factor $g(z)$ is given by:
\begin{equation}
g(z)=1-8z^2+8z^6-z^8-24z^4\log z,
\end{equation}
here we use $m_c/m_b=0.316$.
If we take experimental result $BR(B \to
X_c e\overline{\nu} ) =10.8\% $~\cite{data}, the branching ratios of
$B \to X_s \gamma$ is found.

The effects of QCD corrections from $m_t$ to $M_W$ can first be
seen from values of $C_i(M_W)$. The Figure 5 in previous paper~\cite{lcd1}
shows that the effects of the QCD corrections are roughly within
ten percent and not depend on $\tan \beta$ very much. However,
one will see soon that the effects, though only in ten percent,
will make substantial changes for
the constraints on the parameter space of 2HDM.

Applying the CLEO newer experiment of $b\to s \gamma$ decay,
$1.0\times 10^{-4} < Br( B\to X_s \gamma) < 4.2\times 10^{-4}$, at 95\%
C.L.~\cite{cleo2}. One can obtain the excluded region for model parameter
$\tan \beta$ and $m_{\phi}$. Without QCD corrections from $m_t$ to
$M_W$, the excluded region is more sensitive for changing of
$\alpha_s$ than for $m_t$~\cite{s1}. But after including this QCD
corrections, it is interesting to note that the parameter space is more
sensitive for changing of $m_t$ than for $\alpha_s$, especially in
Model II.~\cite{lcd3}

for Model I, there are two bands
in the $\tan \beta$-$M_\phi$ plane, excluded by our reanalysis with
the latest measurements on $b\to s\gamma$ and $m_t$~\cite{lcd3}.
The excluded region is large, only two narrow windows in the parameter
space are allowed.
For model II, the analysis shows that there is a lower bound for
mass of charged Higgs: 310 GeV, at 95\%C.L.. The previous analysis~\cite{s1}
without QCD corrections from $m_t$ to $M_W$ gave a bound only 200 Gev.
So the bound is more strict now.

In conclusion, due to the QCD corrections from $m_t$ to $M_W$,
the new experimental value of $m_t$ and the bounds for
$b \to s \gamma$, the constraints for 2HDM are strained substantially.
For instance, the lower bound for the mass of the charged Higgs is put
up at least $100 GeV$ for Model II.

\vspace{0.5cm}
\noindent{\bf Acknowledgements}

The author would like to thank Prof. Z.X. Zhang for collaboration.

\end{document}